\documentclass{vgtc}           
\ifpdf
  \pdfoutput=1\relax                   
  \usepackage{graphicx}                
  \DeclareGraphicsExtensions{.pdf,.png,.jpg,.jpeg} 
\else
  \usepackage{graphicx}                
  \DeclareGraphicsExtensions{.eps}     
\fi%

\graphicspath{{figures/}{pictures/}{images/}{./}} 

\usepackage{microtype}                 
\PassOptionsToPackage{warn}{textcomp}  
\usepackage{textcomp}                  
\usepackage{mathptmx}                  
\usepackage{times}                     
\usepackage{cite}                      
\usepackage{tabu}                      
\usepackage{booktabs}                  

\onlineid{0}

\vgtccategory{Research}

\title{Externalizing Transformations of Historical Documents: Opportunities for Provenance-Driven Visualization}

\author{Tomas Vancisin\thanks{e-mail: tv8@st-andrews.ac.uk}\\ %
        \scriptsize University of St Andrews \\ %
\and Mary Orr\thanks{e-mail: mmo@st-andrews.ac.uk}\\ %
     \scriptsize University of St Andrews %
\and Uta Hinrichs\thanks{e-mail: uh3@st-andrews.ac.uk}\\ %
     \parbox{1.4in}{\scriptsize \centering University of St Andrews \\ }}

\abstract{

Transcription, annotation, digitization and/or visualization are common transformations that historical documents such as national records, birth/death registers, university records, letters or books undergo. Reasons for those transformations span from the (physical) protection of the original materials to disclosure of ``hidden'' information or patterns within the documents. Even though such transformations bring new insights and perspectives on the documents, they also modify the documents' content, structure, and/or artifactual form and thus, occlude prior knowledge and interpretation. When it comes to visualization as a means to transform historical documents from written to abstract visual form, there is typically little acknowledgment or even understanding of the previous transformation steps these documents have gone through. The ``tremendous rhetorical force'' \cite{dignazio_data_2020} of visualization, we argue, should not be at the expense of the multiple pasts, contexts, and curators that are inherent in historical record collections. Rather, the urgent question for the fields of visualization and the (digital) humanities is how to better support awareness of these multiple layers of interpretation and the people behind them when representing historical documents. We begin to address this question based on a collection of historical university records by (a) investigating common transformation processes of historical documents, and (b) discussing opportunities and challenges for making such transformations transparent through what we call ``provenance-driven visualization''; the idea for a visualization that makes visible the layers of transformation (including interpretation, re-structuring, and curation) inherent in historical documents.}


\begin{document}
\maketitle

\section{Introduction}
The visualization of historical documents and cultural collections has attracted extensive research within the fields of visualization and the (digital) humanities (see Windhager et al.\cite{windhager_visualization_2018} and J\"anicke et al.~\cite{janicke_analysis_2017} for overviews). Whether the research focuses on building tools that provide new ways for exploration of historical documents~\cite{whitelaw_generous_2015, noauthor_google_nodate}, or on the development of novel visualization techniques tailored for such documents~\cite{hinrichs_emdialog_2008,noauthor_museum_nodate}, the variety and the amount of visualization-focused research shows its far-reaching potential in the area. However, efforts to disclose processes related to data acquisition and document transformation have, so far, focused mostly on the technical steps crucial for visualization \cite{edelstein_historical_2017,hinrichs_trading_2015,hyvonen2017reassembling}. Details on how records were transformed from original into digital/visualization form, and the curatorial decisions that were involved in these transformation steps are typically omitted. Similarly, the resulting interfaces and visualizations often do not allude to the labor and interpretative work involved in these processes. 

Research in visualization and digital humanities has already started to critically discuss the issue of (often) hidden data choices and transformation processes~\cite{Doerk_2013,hullman_visualization_2011,Vancisin_2018}, and their potential societal and political impact~\cite{dignazio_feminist, dignazio_data_2020, correll_ethical_2018}. The question that stands is how to transform theory into practice. How can we characterize these transformation processes and their impact on interpretation of and engagement with historical collections? How can we use visualization to make these processes visible or, at least, more transparent in order to facilitate contextual interpretations and re-engagement with the past knowledge (processes)? Focusing on the case study of the St Andrews Historical University records, an exemplary collection of student records that date back to the 15th century~\cite{1413_maitland,smart_alphabetical,maitland_anderson_matriculation_1905}, we begin to address these questions. The collection is particularly interesting because parts of it have undergone a variety of documented transformations across three centuries (see Fig.~\ref{fig:transformationSteps})---from the original handwritten matriculation rolls to interactive visualizations of the records~\cite{Vancisin_2018}. Through in-depth interviews conducted with experts from the University of St Andrews who have worked with these records at different stages, we capture and characterize these processes and their impact not only on the collection's structure, content, artifactual and representational form, but, ultimately, on the way researchers and the general public can engage with it. Based on this analysis and in the context of previous work in the digital humanities and visualization~\cite{Doerk_2013,hullman_visualization_2011}, we present and define \textit{provenance-driven visualization} as visualizations which focus on disclosing the transformation processes that historical and cultural collections have gone through, e.g., prior to or as part of digitization and visualization processes. Provenance-driven visualization can be considered as a visualization approach and/or visualization-based research method to make such processes explorable. We illustrate the idea of provenance-driven visualization based on a visualization prototype that shows interpretation and transformation work that has been done to the St Andrews Historical University records across three centuries. Part of this visualization work has been discussed in a DH2020 750-word abstract~\cite{vancisin_orr_hinrichs_2020}. Here, we expand on this prior design work by addressing its methodological and theoretical implications. We see provenance-driven visualization as a novel method/perspective to visualization in DH and, ultimately, as an integral part of what we call \textit{digital research ethics}---methodologies and research approaches to data analysis and visualization that focus not only on the content of historical documents and cultural collections, but also on the inherent interpretation and curatorial processes these documents and subsequent data representations embody.

\section{St Andrews' Historical University Records}

\begin{figure}[b!]
    \vspace{-1.2em}
    \centering
    \includegraphics[width=0.8\columnwidth]{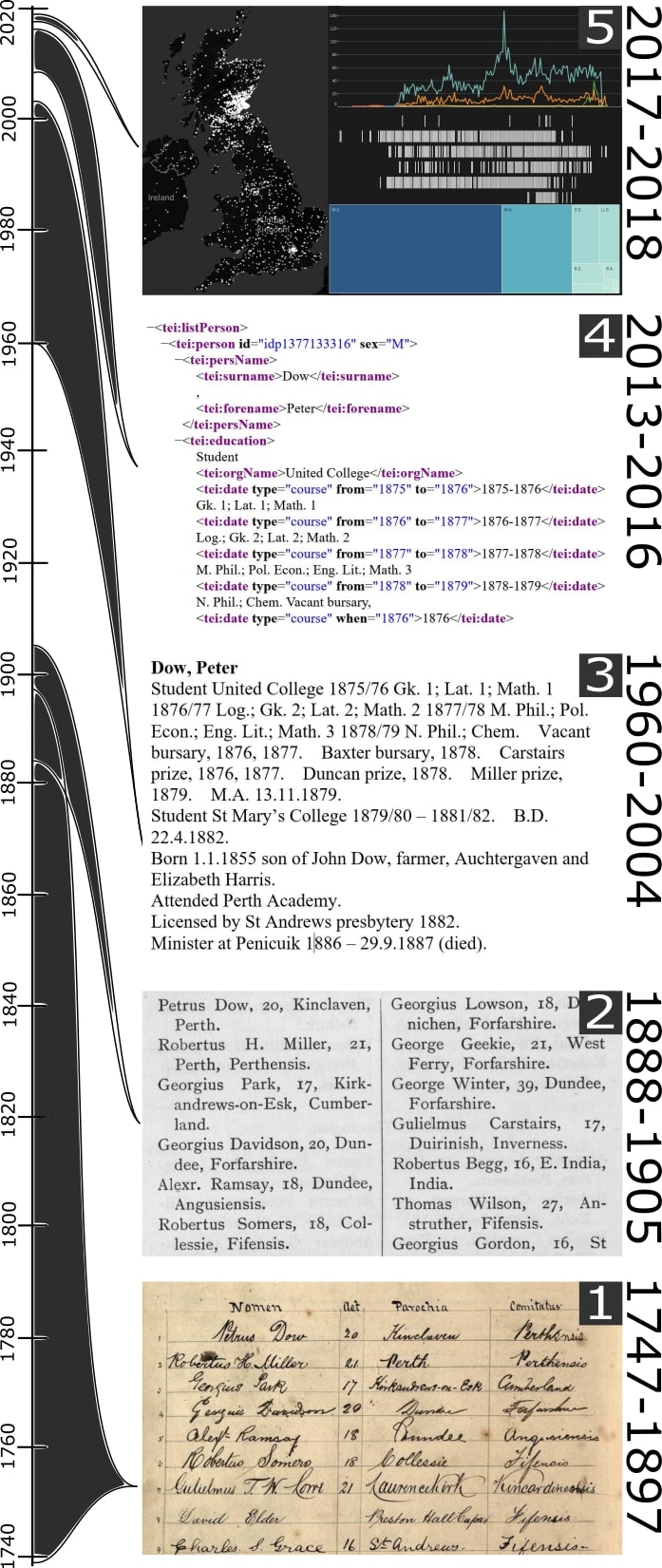}
    \vspace{-1.0em}
    \caption{Transformations of the St Andrews' Historical University Records (1747--1897).}
    \label{fig:transformationSteps}
\end{figure}

The University of St Andrews has been keeping records of its students and staff members since its foundation in 1413. The records provide rich insights into the University's history as well as the societal and political structures at the time. Our case study focuses on the records created between~1747 and~1897 which have undergone a variety of transformations as part of several projects that aimed to preserve and conserve this collection (see Figure~\ref{fig:transformationSteps} for an overview). Originally, each student wrote down their name, and toward the end of this period, also church affiliation and birth place into the Matriculation/Graduation Roll (see Fig.~\ref{fig:transformationSteps}.1). From 1888 to 1905 the then Keeper of Manuscripts and Muniments, James Maitland-Anderson, transcribed these records which resulted in a printed book (see Fig.~\ref{fig:transformationSteps}.2). Anderson's work was re-visited between 1960 and 2004 by another Keeper of Manuscripts and Muniments at St Andrews, Dr Robert Smart, who also modified the records' content, drawing from a large variety of additional sources. He transformed the collection into what is now known as the Biographical Register of the University of St Andrews (BRUSA)~\cite{smart_biographical_2004}, a physically bound alphabetical index of student and staff names that includes information about their demographics, courses taken in St Andrews, parentage, and subsequent careers (see Fig.~\ref{fig:transformationSteps}.3).
Almost 10~years after the publication of BRUSA, the University Library's Digital Humanities and Research Computing team led by Dr. Alice Crawford transformed the register into searchable digital form using the Text Endocing Iniciative (TEI) (see Fig.~\ref{fig:transformationSteps}.4). This resulted in an online interface\footnote{https://arts.st-andrews.ac.uk/biographical-register/} which enables the textual search of BRUSA. In 2018, we transformed the 11,894 TEI files, each representing one person, into a relational database and created a number of interactive visualization sketches in Tableau Desktop that enable the exploration of the records' content from different perspectives~\cite{Vancisin_2018} (see Fig.~\ref{fig:transformationSteps}.5). Our early visualizations take a decidedly quantitative approach to this collection and they have revealed interesting trends and patterns. However, we also found this approach to be potentially misleading as these graphs, charts and maps do not disclose the curatorial and interpretative efforts that have shaped the underlying data. The incompleteness of records, for example, or uncertainty of names and geographic locations become invisible behind definitive graphs and charts~\cite{Vancisin_2018}. This led us to research the history of this collection and the transformation processes it has gone through by conducting interviews with experts who have worked on its previous iterations.

\vspace{-0.3em}
\section{Interviews with Experts}
We conducted seven in-depth interviews with the expert archivists, librarians, historians, and software engineers who have worked with the St Andrews University records at different stages. All participants agreed to be credited with their name for this research. Interviewees included Dr. Robert Smart, the historian, archivist, and paleographer responsible for compiling the records into BRUSA~\cite{smart_biographical_2004} (see Fig.~\ref{fig:transformationSteps}.3). From the team that worked on the transformation of BRUSA into a web-based search platform (see Fig.~\ref{fig:transformationSteps}.4) we interviewed Dr. Alice Crawford (project lead), Siri Hjelsvold, a medieval historian responsible for manually introducing XML:TEI tags to the records, and Patrick McCann and Swithun Crowe, software engineers who helped conceptualizing the tagging framework and implemented the search platform. The interviewees were asked to describe their processes when working with the University records, what motivated these processes, what key decisions and challenges informed their methods, and how they thought the transformations that they applied to the records impacted subsequent interpretations and engagement with this historical collection. While an interview with Maitland-Anderson (1852--1927) about his transcription processes of the original records (see Fig.~\ref{fig:transformationSteps}~\cite{maitland_anderson_matriculation_1905}) was not possible, we interviewed Rachel Hart, senior archivist and the current Keeper of Manuscripts and Muniments at the University who provided rich details about the context in which the original records were created and who also described Maitland-Anderson's working practices. Hart's work on similar document collections also provided invaluable information about the opportunities as well as issues introduced by the digitization of historical records. Finally, we interviewed Sean Rippington, the digital archives officer responsible for curating and implementing the current digital preservation system at the University. While not directly involved with the historical University records, he provided insights into common digitization processes. 

All interviews were transcribed and analyzed for common themes (motivation, process, modifications, challenges, effect, insights, advice, future for the records) using an open-coding and thematic analysis approach~\cite{Boyatzis_1988,Guest_2012}. The qualitative coding focused on characterizing transformation processes and related challenges as well as implications for interpretation. Below, we describe the transformation processes we have identified as part of our interview analysis.

\vspace{-0.3em}
\section{Transformations}
Our qualitative analysis revealed four key categories of transformation processes that the University records have gone through: \textit{Manual Transcription}, \textit{Content Modification}, \textit{Organizational \& Structural Modifications} and \textit{Artifactual \& Representational Form}.

\vspace{-0.3em}
\subsection{Manual Transcription}
The first transformation the records (handwritten, originally in Latin) underwent involved their manual transcription. This process is defined as ``\textit{the effort to report---insofar as typography allows---precisely what the textual inscription of a manuscript consists of.}''\cite[p.201]{meulen_system_1999}. The records were transcribed by Maitland-Anderson and later by Smart, who verified and in some cases re-transcribed Maitland-Anderson's work. Our interviews with Hart and Smart highlight the effort and level of interpretation inherent in this process which involves extensive experience in paleography: ``\textit{[...] it takes time, it takes experience, and you have to learn how to read the old hands.}''~[Hart]. A paleographer also often needs to transcribe Latin texts: ``\textit{Latin has a lot of abbreviations within it, so, immediately, you need to have somebody who can understand Latin and expand abbreviations correctly.}''~[Hart]. Hart also stresses that a transcription of historical records can never be considered a reproduction of the originals; interpretation is necessary: ``\textit{You're dependent on the ability to read the language but also to read the hands in order to be able to interpret. And this is why it's never a 100\% certain that the person who's transcribing has got it absolutely right.}''~[Hart]. Smart himself acknowledges this in relation to transcription: \textit{``The further back in time you get, the more difficult it becomes, so that with the present one [student records from 1413 to 1579], I am not even sure if I got the names right.}''~[Smart]. Interpretation is necessary in the transcription process and it will introduce uncertainties, but without the meticulous work of Anderson and Smart, the knowledge within the original Roll would only be available to paleographers: and by transcribing the Roll, they have protected the physicality of the original materials and its onward curation.

\vspace{-0.3em}
\subsection{Content Modification}
Maitland-Anderson aimed to preserve the content included in the original Matriculation/Graduation Roll. Smart however, deliberately excluded some student information such as their age~\cite{smart_biographical_2004}. At the same time, he vastly expanded the demographic information about students and staff by researching the University archives (e.g., library records, class lists, or medical degree testimonials) as well as national and church records, academic publications, newspapers, individual/family/national biographies, and history books for additional information. He even corresponded with living relatives and traveled to graveyards to find information on monumental inscriptions. As part of his archival work, Smart had to interpret information from multiple record collections in order to extract usable and consistent snippets to include in the existing student and staff records. His curatorial expansion of the historical records is remarkable and provides a much richer picture of University students and staff than the original records. However, Smart himself also emphasizes the limitations of his work in terms of completeness: ``\textit{I simply used the sources that were available at the time. But since it [BRUSA] was published, of course, a lot of new resources have become available. The Internet has become available. I didn't have any of that.}''~[Smart]. Crawford's project further expanded the historical records by adding URLs to student and staff publications where available. While all these expansions of the records were done manually through extensive research, we expanded the records computationally using Google's geocoding API to link locations of birth and death with exact geographic coordinates---a requirement for the creation of geospatial visualizations that introduce uncertainties due to ambiguities in historical place names~\cite{Vancisin_2018}. These expansions of the original records have contributed to the records' overall value and research potential, but also, again, introduced additional interpretation layers as well as uncertainty.

\vspace{-0.3em}
\subsection{Organizational \& Structural Modifications}
Another category of transformation processes includes organizational and structural changes which can have a strong influence on how people engage with and make sense of historical and cultural collections. Maitland-Anderson decidedly aimed to avoid modifications of the original records as much as possible: ``\textit{The reader of the printed Roll is, thus, as nearly as may be, in the same position as the consulter of the manuscript Roll.}''\cite[p.62]{maitland_anderson_matriculation_1905}. However, his transcribed version of the records moves away from the original records' tabular representation by excluding the explicit labeling of individual parts of the records (see Fig.~\ref{fig:transformationSteps}.1 \& 2). He also removed the numbering of individual records. Nevertheless, the order of listed records still mirrors the order in which students signed the Matriculation Roll. A more major structural modification is introduced by Smart who moved away from this originally temporal structure of the records and organized them alphabetically. This enables easy look-up of individual names, but the inherent chronological order of the records is lost. Smart also introduced an implicit internal structure to the additional information he gathered for each record. All records contain consistent sections (name, education, birth, floruit, and death), although these are only visible in each record's internal structure; no explicit labels are provided. 

This internal structure was kept and further emphasized in Crawford's project where consistent tags ($<$name$>$, $<$education$>$, $<$birth$>$, $<$floruit$>$ and $<$death$>$) were applied to each record. Tagging makes the implied internal structure of each record explicit and allows the identification of individual record parts across the collection. The process requires an interpretative effort, as Patrick McCann explains: ``\textit{[With TEI] you got a very rigid structure. [Rigid] in terms of the kinds of elements you can have and the kinds of information those things can describe. So, there is necessarily a change to the data in that process.}''. When it comes to the external structure of the records, Crawford's team divided the register into 11,894 individual XML:TEI files without any order. The order and organization of records purely depends on search queries put forward by the user. For example, text-searching for a particular student name will bring up all records that contain this name. Our process of transforming the TEI-tagged records into a relational database further emphasizes the rigid structure introduced by Crawford and colleagues (based on Smart's prior work): information included in each record is segmented into tables. This enables more flexibility when it comes to searching and visually representing the records, but can be considered as a strong interpretation step that permits certain perspectives on the records and hinders others.

\vspace{-0.3em}
\subsection{Artifactual \& Representational Form}
\label{artifactualForm}
Figure~\ref{fig:transformationSteps} clearly shows that as part of curatorial and interpretation processes, the records have fundamentally changed both in their artifactual and representational form. Hart describes the original Matriculation/Graduation Roll as a ``\textit{[...] lovely big book. Physically large, it's labeled Matriculation and Graduation Roll 1739 - 1888. [...] This is clever, because it has one end---matriculations---and on another end, backwards, it has graduations. They've used the same volume for two purposes, and they've simply turned it over in the middle.}'' Maitland-Anderson describes the Roll as ``\textit{an autograph album of a most interesting kind.}''~\cite[p.62]{maitland_anderson_matriculation_1905}. The ``mechanical print'' artifacts produced by Maitland-Anderson~\cite{maitland_anderson_matriculation_1905} and Smart~\cite{smart_biographical_2004} broaden this collection's audience to non-expert readers, but remove further paleographical inquiry, and do not support the same kind of intriguing reading affordance or human ``touch'' implied by the hand-crafted book and handwritten text. 
\begin{figure*}[h!]
    \centering
    \includegraphics[width=2.0\columnwidth]{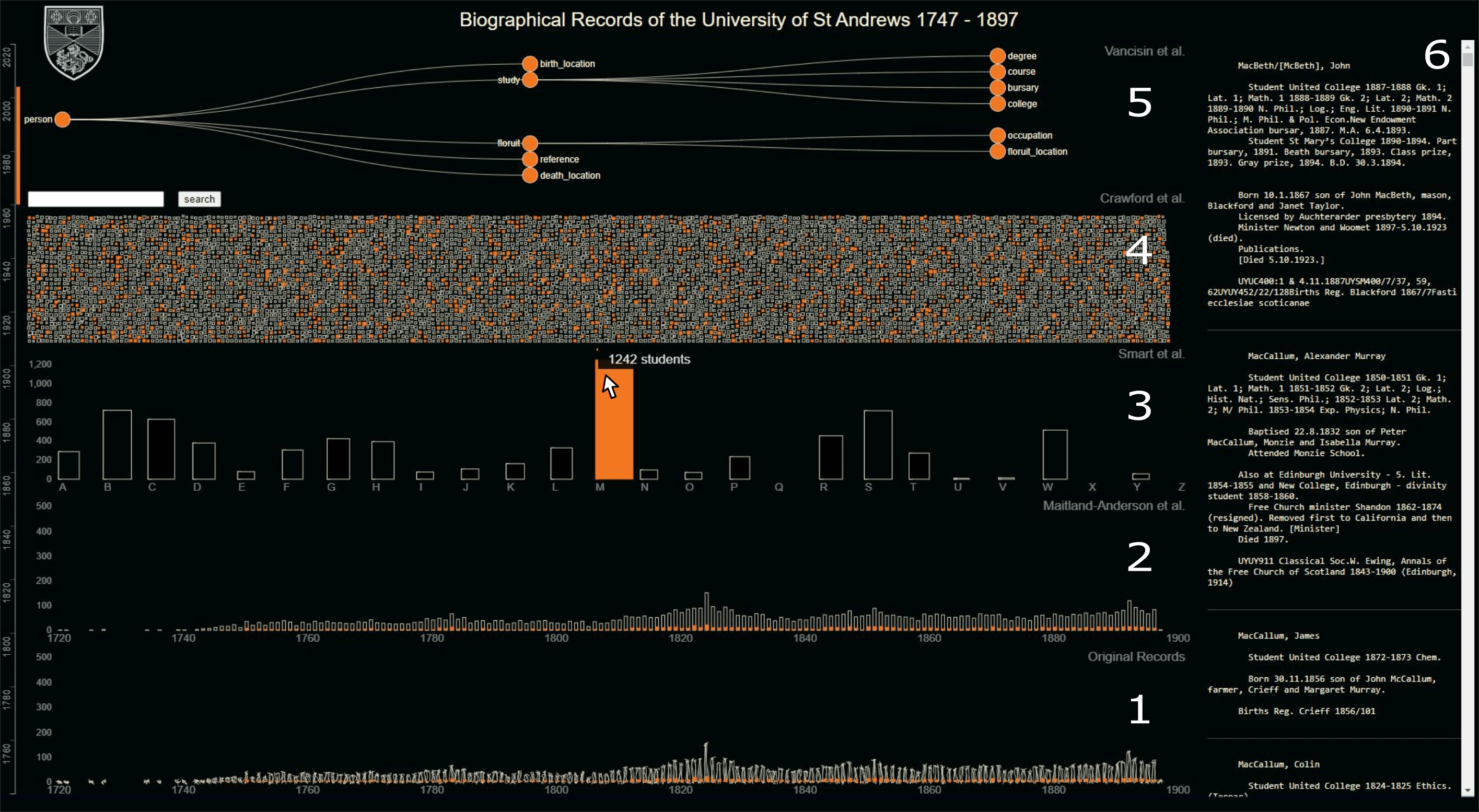}
    \vspace{-1.0em}
    \caption{Prototype to illustrate what a provenance-driven visualization could look like.}
    \label{fig:process}
    \vspace{-1.2em}
\end{figure*}

The transformation of the records from physical to digital form is an even more significant change. Physical affordance is replaced by on-screen interactions; nothing remains of the aesthetics and materiality of the original records---a potential problem discussed in previous research~\cite{Forlini_2018}. The University records are a good example of this: their original artifactual form illustrates, for example, the diversity of individuals attending the University at the time, as visible in their unique signatures~\cite{Vancisin_2018}, and the ways in which knowledge was passed on at the time through record keeping. Although searchable by name and keyword, the digital versions of the records occlude all prior efforts of transcription and interpretation in a way that also changes how the information is represented artifactually and epistemologically (e.g., alphabetical re-ordering). 

Apart from the visualization, all modification processes applied to the University records are text-based. Our transition from text to abstract visual representations is perhaps the most (visually) notable transformation the records have gone through. For example, birth locations are given geographical context though a map view, and people from the same country of birth are aggregated in a bar chart view to provide an overview of students' demography. This visual transformation of records enables ``birds-eye'' perspectives on the records and reveals high-level trends and patterns. However, individual records and the people behind them get lost along with all the previous transformations. The viewer is left with a mere tip of an information/context ``iceberg'' which, even though rhetorically powerful, can only portray certain perspectives on the records.

It is crucial to emphasize that while these categories apply to our collection of historical University records, other types of historical and cultural collections may reveal additional categories of transformations and also additional nuances to the ones introduced here. However, we do believe that the list we present here, in some form, likely applies to other historical and cultural collections. What our case study shows is that there is not only a variety of transformation processes that take place when digitizing historical collections, but also that these processes have a strong influence on (1)~the content of records, (2)~how these records, individually and as a collection, can be represented, and (3)~how people may engage with and interpret them on a physical and cognitive level. Representing only the final stage of data collected from such collection---as is most often the case---deprives the viewer of important contextual information and, therefore, can skew interpretation. Again, our own initial visualizations (see Fig. 1.5) may lead viewers to assume completeness of the student and staff records from 1747--1897 when there are not only gaps, but also layers of interpretation and curatorial decisions. This is also problematic from an ethical perspective because the people involved in these transformations often remain unacknowledged~\cite{correll_ethical_2018,dignazio_data_2020}. This led us to the question of \textit{if} and \textit{how} visualization could be leveraged to not only focus on historical and cultural collections' content, but also on the transformation processes that such collections have gone through and the people behind these processes.

\section{Provenance-Driven Visualization}
The idea of provenance-driven visualization is in line with recent discussions in the field of visualization and the (digital) humanities toward critical approaches to data- and visualization-driven research processes. Diakopulous \& Hullman  emphasize the importance of data provenance in narrative visualization as a means for ``transparency and trustworthiness of the presentation source to the end-users''~\cite[p.2234]{hullman_visualization_2011}. A similar approach is pointed at by Doerk et al. who also highlight the importance of trustworthiness in the field of visualization that can be achieved through disclosure of decisions made with the data~\cite{Doerk_2013}. Drucker's critical reflection on the role of visualization in humanities research stresses the constructed nature of data and calls for visualization approaches that portray data as ``interpreted knowledge, situated and partial, rather than complete''~\cite{drucker_humanities_2011}. Correll highlights the importance of considering the ethical implications of visualization from the perspective of giving proper credit to the people and labor involved in all related processes~\cite{correll_ethical_2018}.

On a practical level, in order to provide  context for a visualization and its underlying data, Wrisley has introduced the concept of \textit{pre-visualization}~\cite{wrisley_pre-visualization_2018}. This idea aims to introduce textual prefaces that are common in books (see Maitland-Anderson~\cite{maitland_anderson_matriculation_1905} or Smart~\cite{smart_biographical_2004} as examples), or in web-based projects' `About' sections (see Crawford\footnote{https://arts.st-andrews.ac.uk/biographical-register/about-the-project/}). Another approach is introduced by Peoux \& Houllier who propose abstract process diagrams to reveal transformation processes in the context of information management~\cite{peoux_visualize_2017}. In contrast, provenance-driven visualization takes a data/visualization-driven approach where transformation processes inherent in a historical or cultural collection are not only identified, but also characterized in the form of data and then visually represented in the form of static or interactive visualizations so they can be directly explored by viewers, perhaps even alongside more content-based visualizations. As such, provenance-driven visualization could be considered as a visual and interactive type of pre-visualization. Provenance-driven visualization provides a visual trace of the multiple contexts, formats, and curatorial decisions embodied in historical data, recognizing the importance of the onward value and interpretation of such decisions. Below we illustrate the idea of provenance-driven visualization, based on our case study of the St Andrews' Historical University Records.

\section{Provenance-driven Visualization Prototype}
Figure~\ref{fig:process} shows an example of a provenance-driven visualization\footnote{https://tv8.host.cs.st-andrews.ac.uk/provenanceDrivenVisualization/} that we designed to make visible many of the transformation processes inherent in the St Andrews' historical University records. The visualization highlights the stages of transformation in the form of five layers (see Fig.~\ref{fig:process}.1--5). Each layer represents the characteristics of the historical records as a result of the transformation processes they have gone through. The visualization combines overviews of all records to highlight changes in their content and structure (see Fig.~\ref{fig:process}.1--5) with details on individual records enabling a close-reading perspective (see Fig.~\ref{fig:process}.6).

\begin{figure}[b!]
    \vspace{-1.2em}
    \centering
    \includegraphics[width=0.8\columnwidth]{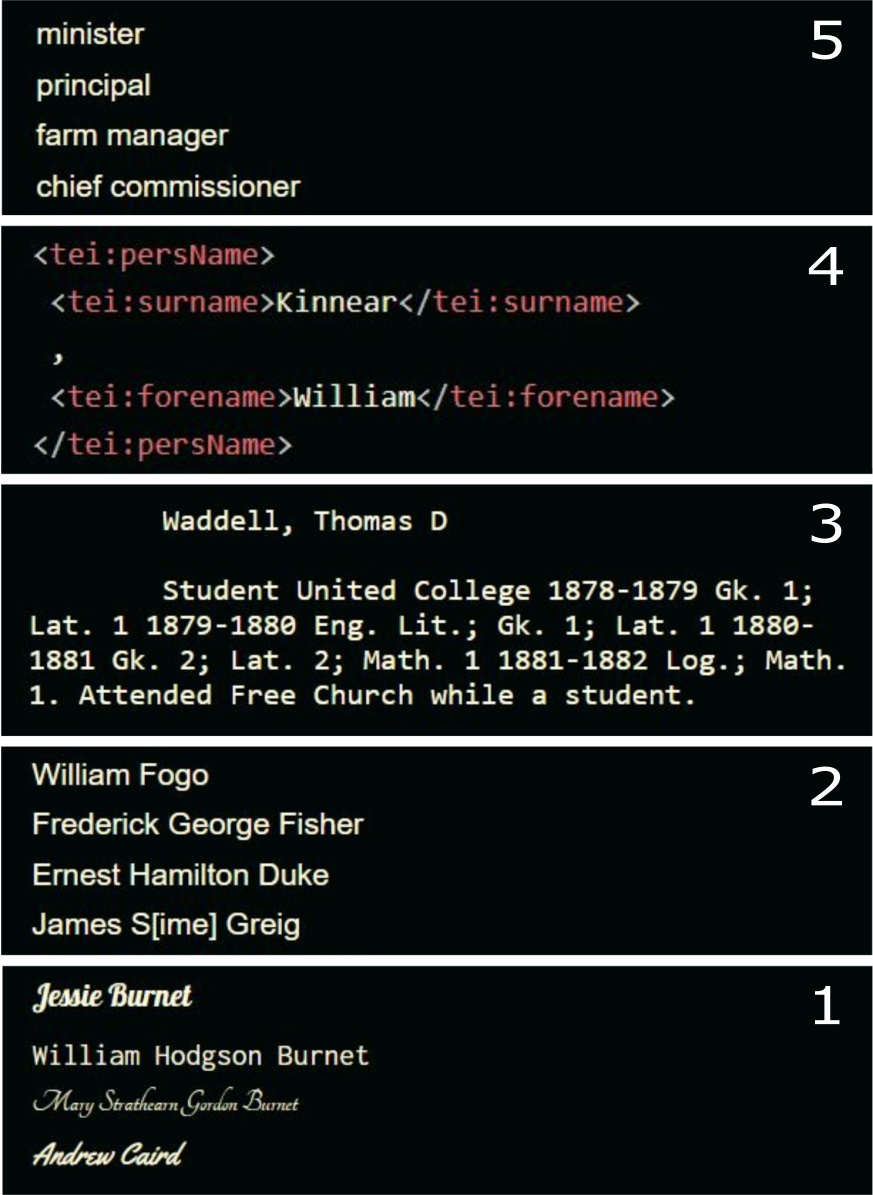}
    \vspace{-1.0em}
    \caption{Internal Transformations.}
    \label{fig:text}
\end{figure}

\paragraph{Layer~1: Original Records.} The amount and structure of original records is represented in form of a temporal bar chart at the bottom layer of the visualization (see Fig.~\ref{fig:process}.1). Student records are aggregated and organized according to their temporal distribution, hinting at the chronological order in which student signatures were initially collected in the Matriculation Roll. Using a sketch-based stroke for bars\footnote{https://github.com/sebastian-meier/d3.sketchy/blob/master/README.md} we emphasize the unique characteristics of the original, handwritten records' \textit{Artifactual \& Representational Form}. Hovering over a bar reveals the student numbers in this year and provides individual names of corresponding students to the right in the ``Record View'' (see Fig.~\ref{fig:text}.1).  Variation in fonts emphasizes the `different hands' that signed the Matriculation Roll.

\paragraph{Layer~2: First Transcription.} Maitland-Anderson's transcription work on the records left the temporal distribution of student records unchanged; so we represent student records, again in form of a temporal bar chart in this layer (see Fig.~\ref{fig:process}.2). However, to emphasize the results of Maitland-Anderson's transcription process that transformed the hand-written records into print form we used a smooth stroke for the bar chart, hinting at the unifying effect of this process. As in the previous layer, hovering over a bar reveals the student numbers in this year and provides individual names of corresponding students to the right (see Fig.~\ref{fig:text}.2). However, student names are shown in the same font, again, hinting at the print-based character of records after transcription.

\paragraph{Layer~3: Expansion \& Re-structuring.} The next layer shows the result of Smart's work on the student records, highlighting in particular his \textit{Content Modification} (expansion of record content) and \textit{Organizational \& Structural Modification} (from chronological order to alphabetical index) (see Fig.~\ref{fig:process}.3). The bar chart remains as the visualization technique of choice, but records are aggregated by the first letter of students' last names, and bar width represents the aggregated amount of content (word count) by alphabetical index. We keep the depiction of bars in smooth lines, again, hinting at the print-based format of the records at this stage. Hovering over a bar reveals corresponding student numbers and shows corresponding student records to the right (see Fig.~\ref{fig:text}.3). Here, Smart's expansion and structuring of individual records becomes visible as student names are expanded with additional demographic information. 

\paragraph{Layer~4: Re-Organization \& Re-structuring via TEI.} Crawford's work revoked both the temporal and alphabetical structure of the historical University records and applied more structure to individual records. Our visualization reflects this by portraying each individual record as a square where squares are randomly arranged into a pile with no inherent ordering (see Fig.~\ref{fig:process}.4). Hovering over a square reveals the corresponding student record in TEI form to the right (see Fig.~\ref{fig:text}.4), including the applied TEI tags that provide a stronger structure to each record. 

\paragraph{Layer~5: Structuring into a Relational Database.} The most recent transformation process---the \textit{Organizational \& Structural Modification} of Crawford's TEI files into a relational database is represented in the top-most layer of our provenance-driven visualization. The database structure is represented as a hierarchical tree diagram where each circle represents a table and links depict relations between them (see Fig.~\ref{fig:process}.5). This representation emphasizes how records are no longer treated as individual entities; their content has been broken down into segments (e.g., locations of birth and death, degree, college, occupation, etc.). Hovering over one of the nodes will reveal data that can be extracted from the corresponding table (e.g., occupations, see Fig.~\ref{fig:text}.5).

Key to our visualization is that all layers are interactive and interlinked. Hovering over an element in one layer automatically brings up the corresponding records in the ``Record View'' to the right and also highlights these in the other visualization layers. For example, when hovering over a year bar in the original record layer at the bottom, the same year is highlighted in Maitland-Anderson's layer, and all student records from that year are shown in Smart's layer, according to their last names. In Crawford's layer individual squares corresponding to these records are highlighted in orange and in our database representation, the user can see which tables in the database contain which parts of the records (see Fig.~\ref{fig:process}). 

We consider this visualization prototype as one example of a provenance-driven visualization---of course other approaches, also incorporating different visualization techniques are possible. We found that organizational and structural transformations can be visualized relatively easily by modifying the grouping and/or spatial position of visual elements. Other aspects such as changes in \textit{Artifactual \& Representational Form} are more difficult to depict. For example, in our visualization the material form of the original records or subsequent books produced by Maitland-Anderson and Smart are still invisible---we would like to incorporate this in from of scanned snippets of original records, but this is not without significant effort in terms of data preparation. 

We also highlight that a provenance-driven visualization as presented here should not be considered as a replacement of textual background information of the collection and corresponding data represented---some textual explanations of curatorial decisions and transformations are likely necessary and should be incorporated into the visualization. However, we believe that a visual and interactive form of representing the often layered transformation processes of historical and cultural collections can be evocative and raise curiosity that may promote critical perspectives on the visualization of historical and cultural collections, stimulate discussion, and, ultimately, point the viewer to engage in more in-depth research on the history and background of such collections. Based on our interviews and the case study presented here, we discuss opportunities for provenance-driven visualization, also in the context of ongoing work at the intersection of (digital) humanities and data visualization.


\vspace{0.2em}
\section{Discussion}
We see the concept of provenance-driven visualization as an opportunity to (1) promote \textit{attribution \& fairness} by acknowledging the laborious process of making historical and cultural collections available to broader audiences, (2) expose \textit{different layers of knowledge and interpretation} that may have changed throughout the years, (3) promote \textit{transparency} of transformation processes that have been applied to the collection, and (4) encourage \textit{interdisciplinary research}.

\paragraph{Attribution \& Fairness.}
Working with historical and cultural collections in order to preserve them and/or to make them available to a broader audience is a huge effort. Our interviews revealed years of work behind each transformation of the historical University records. For example, Smart worked on the expansion of the records for 60~years. 
Making transformations of historical records visible ensures that this type of labor and the people involved are properly acknowledged. Provenance-driven visualization provides an opportunity here as it can help to disclose the effort and nuances in this work in evocative ways.
This approach is in line with feminist approaches to data analysis and visualization~\cite{dignazio_feminist,dignazio_data_2020} as introduced by D'Ignazio and Klein who advocate for a stronger acknowledgment of the hidden labor involved in data-driven analysis processes. Provenance-driven visualization can thus be considered a practical approach to ensure ethical practices in visualization as advocated by Correll who argues that we ``\textit{[...]ought to visualize hidden labor. Properly acknowledging and rewarding people for their labor is a key component of fairness. Certain kinds of labor (especially those performed by marginalized groups) are under-represented or under-valued in our current schemes of commodification or valuation.}''\cite[p.8]{correll_ethical_2018}. Our provenance-driven visualization prototype addresses this by making visible the transformation layers alongside people's names responsible for these. However, future work should investigate other forms of provenance-driven visualization that put, for example, the historians, archivists, librarians, paleographers, and computer scientists involved in such processes even more into focus.



\paragraph{Exposing Different Layers of Knowledge-Making.}
Drucker argues, that the ``\textit{history of knowledge is the history of forms of expression of knowledge, and those forms change. What can be said, expressed, represented in any era is distinct from that of any other, with all the attendant caveats and reservations that attend to the study of the sequence of human intellectual events, keeping us from any assertion of progress while noting the facts of change and transformation.}''~\cite{drucker_humanities_2011}. Engaging in provenance-driven visualization is an opportunity for exploring and exposing transformations of knowledge expression. Visualizing the transformation of historical record collections can be considered an open-ended inquiry into their contents. This means representing knowledge not as absolute but rather as layered, organic and ever-evolving. Maitland-Anderson's transcriptions of the University records, for example, not only made available the information `hidden' behind the many student hands to a wider audience, but his work is also a manifestation of practices and technologies at the time: he followed archiving practices common for the era, and his curatorial decisions would no be the same hundred years before or after his work. Taking records' transformations into consideration through provenance-driven visualization is an opportunity to highlight changes in knowledge production practices and underlying assumptions. However, this also raises the question of whether there are ways of integrating or combining provenance-driven with traditional, content-focused approaches to visualization that aim at representing selected perspectives on the collection based on corresponding data in its final stages.


\paragraph{Promoting Transparency.} 
Transparency about data, design and research processes has been highlighted as key for visualization design studies~\cite{meyer_criteria_2019}. 
To achieve this, textual descriptions and/or diagrams are often used to portray the transformation processes involved in preparing a collection for visualization~\cite{hinrichs_trading_2015,peoux_visualize_2017}. With provenance-driven visualization we argue for a more data-driven and visual approach to making transformation processes of records visible and explorable in order to allow for a better understanding of curatorial decisions and their impact on the data and subsequent interpretations. We see this as an opportunity for humanities and visualization researchers to engage with and reflect on previous work conducted on the collection at hand in order to inform subsequent data processing and design approaches. Moreover, it is also an opportunity for the general public to better understand the background of historical or cultural collections represented in digital space. However, the concrete impact of provenance-driven visualizations to promote the critical interpretation of historical or cultural collections has yet to be studied in detail, and we invite researchers in the  humanities and visualization to actively engage in this endeavor. 

\paragraph{Encouraging Interdisciplinary Research.}
Provenance-driven visualization and, with it, the visual disclosure of different stages of transformation and interpretation of historical and cultural collections can also be an opportunity for encouraging interdisciplinary research not only at the intersection of visualization and humanities fields, but also involving public audiences. Many researchers from the humanities and visualization community have pushed for such an approach ~\cite{hinrichs_2019,hall_2020} and also for incorporating more diverse perspectives in visualization research~\cite{lee_broadening_2019}. Through provenance-driven visualization we aim to trigger new questions about historical and cultural collections, both regarding insights they can promote as well as opportunities for design. The different perspectives that archivists, paleographers, digitization officers, visualization researchers, and data analysts have about such collections can be very enriching, and we believe that both the design and exploration of provenance-driven visualizations can promote interesting discussions. 

\section{Conclusion}
We have introduced and illustrated the concept of provenance-driven visualization as an approach to visualizing historical and cultural collections that focuses on the explorations of records through the lens of the layered transformation processes they have gone through. By externalizing the ``tremendous rhetorical force''\cite{dignazio_data_2020} of these layers, provenance-driven visualization exposes the individual and combined curatorial and interpretative efforts of the people who have been working with these collections. In contrast to visualization approaches that focus on representing the content of historical and cultural collections in its final processed stage, provenance-driven visualization enables viewers to see a more nuanced perspective on the curatorial history of such collections in order to inform critical interpretation and research perspectives. Our case study have outlined opportunities for provenance-driven visualization which gives credit to the people and labor involved in preparing historical and cultural collection for digitization and visualization, exposes different layers of knowledge-making, promotes data transparency and underlying curation processes, and, ultimately, encourages interdisciplinary research. We hope this paper will spark further practical explorations of provenance-driven visualizations and how this approach can impact the interpretation of historical and cultural collections. 

\bibliographystyle{abbrv-doi}
\bibliography{references}
\end{document}